\documentstyle[11pt]{article}

\begin{document}
\title{\bf High dimensional Schwarzian derivatives and Painlev\'e integrable models}

\author{Sen-yue Lou$^{1,2,3}$\thanks{Email:
sylou@mail.sjtu.edu.cn}, Shun-li Zhang$^2$ and Xiao-yan Tang$^{2,3}$\\
 \it \footnotesize \it $^{1}$CCAST (World
Laboratory), PO Box
 8730, Beijing 100080, P. R. China\\
\it \footnotesize \it $^{2}$Physics Department of Shanghai Jiao
Tong University,
Shanghai 200030, P. R. China\\
\footnotesize \it $^{3}$Abdus Salam International Centre for
Theoretical Physics, Trieste, Italy}

\date{}

\maketitle

\begin{abstract}
Because of all the known integrable models possess Schwarzian
forms with M\"obious transformation invariance, it may be one of
the best way to find new integrable models starting from some
suitable M\"obious transformation invariant equations. In this
paper, the truncated Painlev\'e analysis is used to find high
dimensional Schwarzian derivatives. Especially, a three dimensional Schwarzian
derivative is obtained explicitly. The higher dimensional higher
order Schwarzian derivatives which are invariant under the
M\"obious transformations may be expressed by means of the lower
dimensional lower order ones. All the known Schwarzian derivatives
can be used to construct high dimensional Painlev\'e integrable
models.
\end{abstract}

\vskip.1in

\section{Introduction}

Soliton as the most basic excitation of the integrable models has
been widely applied in natural science$\cite{Nat}$. However, most
of the present studies of the soliton theory and soliton
applications are restricted in (1+1)- and (2+1)-dimensions. The
essential reason is the lack of known higher dimensional
integrable systems.

Actually, to our knowledge, almost all the known integrable (1+1)-
and (2+1)-dimensional models possess Schwarz invariant forms which
are invariant under the M\"obious transformation (conformal
invariance)$\cite{Weiss, Nucci, Lou1}$. Starting from the
conformal invariance of a known integrable model, one may obtain
various other interesting integrable properties. For instance, the
conformal invariance of the well known Schwarzian Korteweg
de-Vries (SKdV) equation is related to the infinitely many
symmetries of the usual KdV equation$\cite{SchKdV}$. The Darboux
transformation$\cite{DT}$ and the B\"acklund
transformation$\cite{Weiss}$ can be obtained from the conformal
invariance. In Ref. $\cite{Lou2}$, it is pointed out that every
(1+1)-dimensional M\"obious invariant model possesses a second
order Lax pair. The flow equation related to the conformal
invariance of the SKdV equation is linked with some types of
(1+1)-dimensional and (2+1)-dimensional sinh-Gordon (ShG)
equations and Mikhailov-Dodd-Bullough (MDB)
equations$\cite{Riccati}$.

According to the above marvelous properties of the Schwarzian
forms, in $\cite{Lou1}$, one of the present authors (Lou) proposed
that starting from a conformal invariant form may be one of the
best ways to find integrable models especially in high dimensions.
Some types of quite general Schwarzian equations are proved to be
Painlev\'e integrable. In $\cite{PRL}$, Conte's conformal
invariant Painlev\'e analysis$\cite{Conte}$ is extended to obtain
high dimensional Painlev\'e integrable Schwarzian equations
systematically. Some types of physically important high
dimensional nonintegrable models can be solved approximately via
some high dimensional Painlev\'e integrable Schwarzian
equations$\cite{Ruan}$.

In all our previous papers, we used only one dimensional Schwarzian
derivative to find some possible new integrable models especially
in high dimensions. Now two important and interesting questions
are: Are there any high dimensional Schwarzian derivatives (high
dimensional derivatives with M\"obious transformation invariance)
and can we obtain some new types of integrable models by using
high dimensional Schwarzian derivatives if there exist?

In the next section, we establish a possible way to obtain
explicit high dimensional Schwarzian derivatives from the Painlev\'e
analysis of partial differential equations (PDEs). In section 3,
we use the obtained (2+1)-dimensional Schwarzian derivatives to
construct some generalized (2+1)- and (3+1)-dimensional integrable
models. The last section is a short summary and discussion.

\section{Obtain high dimensional Schwarzian derivatives via truncated Painlev\'e analysis}

It is known that the truncated Painlev\'e analysis of the
integrable models will lead to the Schwarzian forms (which are
invariant under the M\"obious transformation) of the original
models. For instance, substituting the truncated Painlev\'e
expansion
\begin{eqnarray}
u=\frac{u_0}{\phi^2}+\frac{u_1}{\phi}+u_2,
\end{eqnarray}
into the Korteweg de Vries (KdV) equation
\begin{eqnarray}
u_t+u_{xxx}+6uu_{x}=0,
\end{eqnarray}
will lead to
\begin{eqnarray}
\sum_{i=0}^6X_i\phi^{i-6}=0,
\end{eqnarray}
where $X_i$ are functions of $\{u_0,\ u_1,\ u_2\}$ and the
derivatives of $\{u_0,\ u_1,\ u_2,\ \phi\}$. To get the Schwarzian
form of the KdV equation, one can simply solve the equations
\begin{eqnarray}
X_i=0,\ i=0,\ 1,\ ...,\ 6.
\end{eqnarray}
For the KdV equation, first three equations ($i=0,\ 1,\ 2$) of (4)
give out the results for $\{u_0,\ u_1,\ u_2\}$ while the fourth
equation ($i=3$) of (4) gives out the Schwarzian form
\begin{eqnarray}
\frac{\phi_t}{\phi_x}+\{\phi;\ x\}+\lambda=0,
\end{eqnarray}
where $\lambda$ is an arbitrary constant related to the spectral
parameter and $\{\phi;\ x\}$ is the usual Schwarzian derivative
\begin{eqnarray}
\{\phi;\ x\}\equiv
\frac{\phi_{xxx}}{\phi_x}-\frac32\frac{\phi_{xx}^2}{\phi_x^2}\equiv
S^{[x]}.
\end{eqnarray}
The remained equations ($i=4,5,6$) of (4) are satisfied identically.

It is also known that the truncated Painlev\'e analysis can
be used to find some exact solutions for non-completely
integrable models$\cite{nonint}$. In $\cite{Lou1}$, it has been
pointed out that starting from a M\"obious transformation
invariant form
\begin{eqnarray}
G(S^{[x_i]},\ C^{[x_jx_k]},\ i,\ j,\ k=1,2,3...)=0,\qquad
C^{[x_jx_k]}\equiv\frac{\phi_{x_j}}{\phi_{x_k}},\
\end{eqnarray}
where $G$ is an arbitrary function of the indicated conformal
invariant quantities, one may obtain various integrable models by
selecting the function $G$ appropriately.
Some concrete forms of $G$ have been given in $\cite{PRL}$.

It is natural that if one can find out more conformal invariant
quantities, then one can include these new conformal invariant
quantities into the function $G$ of (7). So the problem is now how to
find some more explicit independent conformal invariants. In this
section we use simply the truncated Painlev\'e analysis to
find some new conformal invariant quantities especially in high
dimensions.

As in the KdV case, for a concrete nonlinear PDE
\begin{eqnarray}
F(x_i,\ u, \ u_{x_i},\ u_{x_ix_j},\ ...)\equiv F(u)=0,
\end{eqnarray}
no matter whether it is integrable or not, the truncated
Painlev\'e expansion has the form
\begin{eqnarray}
u=\sum_{i=0}^\alpha u_i\phi^{i-\alpha},
\end{eqnarray}
where $\alpha$ is a positive integer. Substituting (9) into (8),
one may have
\begin{eqnarray}
\sum_{i=0}^NX_i\phi^{i-N}=0,\qquad N>\alpha,
\end{eqnarray}
where $X_i$ are functions of $\{u_0,\ u_1,\ ...,\ u_\alpha\}$ and
the derivatives of $\{u_0,\ u_1,\ ...,\ u_\alpha, \phi\}$. Now to
find out possible conformal invariant quantities, we should solve
the equations
\begin{eqnarray}
X_i=0,\ i=0,\ 1,\ ...,\ N.
\end{eqnarray}
$\alpha+1$ equations of (11) should be used to solve out $\{u_0,\
u_1,\ ...,\ u_\alpha\}$. Then substituting the results into the
remained equations may yield some of the conformal invariant
equations and one may find some new conformal invariants.

For concretely, we use the two dimensional elliptic $\phi^4$ model
as a simple example
\begin{eqnarray}
u_{xx}+u_{yy}=\sigma u+\mu u^3.
\end{eqnarray}
Substituting (9) into (12), one can easily find that $\alpha=1$
and $N=3$. The first two equations ($i=0,\ 1$) of (11) fix $u_0$
and $u_1$ as
\begin{eqnarray}
&&u_0=\pm\sqrt{\frac{2}{\mu}(\phi_x^2+\phi_y^2)},\\
&&u_1=\pm\frac16\sqrt{\frac2{\mu}}(\phi_x^2+\phi_y^2)^{-3/2}
(\phi_x^2(\phi_{xx}+\phi_{yy})
+4\phi_x\phi_y\phi_{xy}+f_y^2(3f_{yy}+f_{xx})).
\end{eqnarray}
Substituting (13) and (14) into the third equation ($i=2$) of (11)
yields a Schwarzian form
\begin{eqnarray}
&&2C^{[yx]}(7(C^{[yx]})^2+9)(C_x^{[yx]})^2
+2((C^{[yx]})^2+3)(3(C^{[yx]})^2+1)C^{[yx]}_xC^{[yx]}_y\nonumber\\
&&+C^{[yx]}(5-9(C^{[yx]})^4)(C^{[yx]}_y)^2
-6C^{[yx]}(1+(C^{[yx]})^2)^3(\sigma-S^{[yx]})\nonumber\\
&&+6C^{[yx]}(1+(C^{[yx]})^2)^2((C^{[yx]})^2S^{[xy]}
+C^{[yx]}C^{[yx]}_{yy}+S^{[x]})=0
\end{eqnarray}
which is invariant under the M\"obious transformation, where
\begin{eqnarray}
S^{[xy]}\equiv\frac{\phi_{xxy}}{\phi_y}
-\frac{\phi_{xx}\phi_{xy}}{\phi_x\phi_y}-\frac12\frac{\phi_{xy}^2}{\phi_y^2}.
\end{eqnarray}
The final remained equation ($i=3$) of (11) is not a M\"obious
transformation invariant equation.

From Eq. (15) with (16), we obtain not only the conformal
invariant quantities $C^{[yx]}$ and $S^{[x]}$ but also two more two dimensional
conformal invariant quantities $S^{[xy]}$
and $S^{[yx]}$ which
can also be found in the truncated Painlev\'e analysis of other
two dimensional models like the sine-Gordon
equation$\cite{Weiss2}$. When $y=x$, $S^{[xy]}$ and
$S^{[yx]}$ are both reduced back to the usual Schwarzian
derivative $S^{[x]}$.

To find out possible three dimensional Schwarzian derivatives by using
the truncated Painlev\'e analysis, we should start from some
three dimensional nonlinear PDEs. The following
artificial three dimensional model
\begin{eqnarray}
u_{xyz}-u^4=0
\end{eqnarray}
may be a simplest example to find a three dimensional Schwarzian derivative.

Substituting (9) into (17), one can easily find that $\alpha=1$
and $N=4$. $u_0$ and $u_1$ are fixed by the first two equations of
(11):
\begin{eqnarray}
&&u_0=-(6\phi_x\phi_y\phi_z)^{1/3},\\
&&u_1=\frac{6^{1/3}}{36(\phi_x\phi_y\phi_z)^{5/3}}
(5\phi_z\phi_y^2\phi_x\phi_{xz}+5\phi_x\phi_y\phi_z^2\phi_{xy}
+5\phi_y\phi_z\phi_x^2\phi_{yz}\nonumber\\
&&\qquad
+\phi_y^2\phi_z^2\phi_{xx}+\phi_y^2\phi_x^2\phi_{zz}+\phi_x^2\phi_z^2\phi_{yy}).
\end{eqnarray}
Substituting (18) and (19) into the third equation of (11) yields
a M\"obious transformation invariant equation
\begin{eqnarray}
&&C^{[yx]}\left[12\left(S^{[xy]}+S^{[xz]}\right)\left(C^{[zx]}\right)^4
+\left(12S^{[zx]}+12S^{[zx]}-57\left(C_x^{[zx]}\right)^2\right)
\left(C^{[zx]}\right)^2\right.\nonumber\\
&&\left. -14C_x^{[zx]}C_z^{[zx]}C^{[zx]}-\left(C_z^{[zx]}\right)^2\right]
\left(C^{[yx]}\right)^4+
\left[36\left(2S^{[xyz]}-C_x^{[zx]}C_x^{[yx]}\right)\left(C^{[zx]}\right)^3\right.
\nonumber\\
&&
\left.+\left(12C_z^{[zx]}C_x^{[yx]}-50C_z^{[yx]}C_x^{[zx]}\right)\left(C^{[zx]}\right)^2
-6C^{[zx]}C_z^{[zx]}C_z^{[yx]}\right]\left(C^{[yx]}\right)^3+\left[\left(12S^{[yz]}\right.\right.
\nonumber\\
&&\left.-16\left(C_x^{[yx]}\right)^2+12S^{[yx]}\right)\left(C^{[zx]}\right)^4
+\left(-32C_z^{[yx]}C_x^{[yx]}+6C_y^{[yx]}C_x^{[zx]}\right)\left(C^{[zx]}\right)^3
\nonumber\\
&&
+\left.\left(2C_z^{[zx]}C_y^{[yx]}-9\left(C_z^{[yx]}\right)^2\right)
\left(C^{[zx]}\right)^2\right]\left(C^{[yx]}\right)^2
+\left(-6\left(C^{[zx]}\right)^3C_z^{[yx]}C_y^{[yx]}\right.
\nonumber\\
&&\left. -8\left(C^{[zx]}\right)^4C_y^{[yx]}C_x^{[yx]}\right)C^{[yx]}
-\left(C^{[zx]}\right)^4\left(C_y^{[yx]}\right)^2=0,
\end{eqnarray}
where
\begin{eqnarray}
S^{[xyz]}\equiv
\frac{\phi_{xyz}}{\phi_x}-\frac{\phi_{xx}\phi_{yz}}{\phi_x^2}
-\frac{\phi_{xx}\phi_y\phi_{xz}}{\phi_x^3}-\frac{\phi_{xx}\phi_z\phi_{xy}}{\phi_x^3}
+\frac32\frac{\phi_{xx}^2\phi_z\phi_{y}}{\phi_x^4}
\end{eqnarray}
is a new three dimensional Schwarzian derivative. It is obvious
that when $z=y=x$, the three dimensional Schwarzian derivative
$S^{[xyz]}$ shown by (21) is reduced back to the usual one dimensional Schwarzian
derivative.

In principle, the same procedure can be proceeded further to find
higher order higher dimensional Schwarzian derivatives. For
instance, use the truncated Painlev\'e analysis to the
following equation
\begin{eqnarray}
u_{xyzt}=u^3,
\end{eqnarray}
we get a complicated four dimensional fourth order Schwarzian
derivative which is a generalization of the usual fourth order one
dimensional Schwarzian derivative${\cite{higher}}$. It is known that in one
dimensional case, higher order Schwarzian derivative can be
calculated out from lower order Schwarzian derivative $S^{[x]}$
${\cite{higher}}$. The similar situation occurs for high
dimensional high order Schwarzian derivatives. The four
dimensional fourth order Schwarzian derivative obtained from the truncated
Painlev\'e expansion of (22) can also
 be expressed by the lower dimensional
lower order Schwarzian
 derivatives $S^{[xyz]},\ S^{[xy]},\ S^{[xz]},\
S^{[yz]},\ S^{[x]}$ and $C^{[xy]}$ etc. So here we do not write down any
complicated higher order
higher dimensional Schwarzian derivatives.

\section{New high dimensional Painlev\'e integrable Schwarzian
 equations}

In this section we are interested in whether
the obtained new high dimensional Schwarzian derivatives can be
used to construct new high dimensional integrable models. It is quite
fortunate for us to obtain various new higher
dimensional integrable models by using the high dimensional Schwarzian derivatives
obtained from the last section.  Here we list only some simple
special examples on the Schwarzian KdV type extensions.

\subsection{(2+1)-dimensional Schwarzian KdV equation with two
dimensional\\ Schwarzian derivatives}

Usually, it is difficult to obtain isotropic higher dimensional integrable extension(s) of a known lower
dimensional integrable model. However, using the conformal invariants, to find some
isotropic higher dimensional integrable extensions becomes a straightforward work. For instance,
\begin{eqnarray}
C^{[tx]}+a_1C^{[ty]}+a_2S^{[x]}+a_3S^{[y]}+a_4S^{[xy]}+a_5S^{[yx]}=0,
\end{eqnarray}
is obviously a (2+1)-dimensional isotropic extension of (1+1)-dimensional
Schwarzian KdV equation.

By using the standard Painlev\'e analysis, we know that Eq. (23) possesses one
expansion around the arbitrary non-characteristic manifold
\begin{eqnarray}
\phi=\sum_{i=0}^\infty \phi_i\psi^{i-1}
\end{eqnarray}
with arbitrary functions $\{\phi_0,\ \phi_1,\ \psi\}$ and the
expansions around arbitrary characteristic manifold
\begin{eqnarray}
&&\phi=f_0(x_1,\ t) +\sum_{i=0}^\infty \phi_i(x_1,\ t)
\left(x_2+\psi(x_1,\ t)\right)^{i+1},
\end{eqnarray}
for arbitrary $\{f_0(x_1,\ t),\ \phi_0(x_1,\ t),\ \phi_1(x_1,\ t),\ \psi(x_1,\ t)\}$, and two
\begin{eqnarray}
\phi=f_0(x_1,\ t)+\sum_{i=0}^\infty \phi_i(x_1,\ t)
(x_2+\psi(x_1,\ t))^{i+3},
\end{eqnarray}
for arbitrary $f_0(x_1,\ t)$ and $\psi(x_1,\ t)$, where $\{x_1,\ x_2\}=\{x,\ y\}$ or
 $\{x_1,\ x_2\}=\{y,\ x\}$. From the expansions (24)--(26),
we know that all the solutions of the
(2+1)-dimensional Schwarzian KdV equation are single valued
about arbitrary non-characteristic
and characteristic manifolds and then the model is completely integrable$\cite{Weiss3}$.

\subsection{(3+1)-dimensional Schwarzian KdV equation with three
dimensional\\ Schwarzian derivative}

Using the three dimensional Schwarzian derivative $S^{[xyz]}$, we can obtain the simplest
(3+1)-dimensional Schwarzian KdV equation
\begin{eqnarray}
C^{[tx]}+S^{[xyz]}=0
\end{eqnarray}
which can be reduced back to the usual (1+1)-dimensional Schwarzian KdV
equation obviously.

Using the standard Painlev\'e analysis one can easily prove that
the (3+1)-dimensional Schwarzian KdV equation (27) is Painlev\'e integrable. Actually, the
Schwarzian KdV equation (27) possesses the non-characteristic
singularities of the form
\begin{eqnarray}
\phi=\sum_{i=0}^\infty \phi_i \psi^{i-1}
\end{eqnarray}
with arbitrary $\phi_0,\ \phi_1, \ \psi$
and the characteristic singularities of the forms
\begin{eqnarray}
&&\phi=f_0(y,z,t)+\sum_{i=0}^\infty \phi_i(y,z,t) (x+\psi(y,z,t))^{i+1},\qquad {
f_0,\ \phi_0,\ \phi_1,\ \psi\ \rm arbitrary},\\
&&\phi=f_0(x,y,t)+\sum_{i=0}^\infty \phi_i(x,y,t) (z+\psi(x,y,t))^{i+1}, {
\qquad f_0,\ \phi_0,\ \phi_1,\ \psi\ \rm arbitrary},\\
&&\phi=f_0(y,z,t)+\sum_{i=0}^\infty \phi_i(y,z,t)
(z+\psi(y,z,t))^{i+3},\qquad f_0,\  {\rm \psi\ arbitrary},
\end{eqnarray}
and
\begin{eqnarray}
\phi=f_0(x,y,t)+\sum_{i=0}^\infty \phi_i(x,y,t)
(z+\psi(x,y,t))^{i+3},\qquad f_0,\ {\rm \psi\ arbitrary}
\end{eqnarray}
which means all the solutions of (27) are single valued about
arbitrary manifold no matter whether the manifold is
non-characteristic or characteristic.

Furthermore, one can prove that the generalized high dimensional
Schwarzian KdV type model
\begin{eqnarray}
\sum_{i=0}^Na_iS^{[x_i]}+\sum_{i=0}^N\sum_{j=0}^Nb_{ij}S_{[x_ix_j]}
+\sum_{i=0}^N\sum_{j=0}^N\sum_{k=1}^Nc_{ijk}S_{[x_ix_jx_k]}+g(C^{[x_ix_j]})=0,
\end{eqnarray}
for some suitable polynomial functions of $C^{[x_ix_j]}$, $g(C^{[x_ix_j]})$,
are Painlev\'e integrable.

\section{Summary and discussions}

In summary, applying the truncated Painlev\'e analysis to high dimensional
high order PDEs, one may obtain some types of high dimensional high
order Schwarzian derivatives. Especially, an explicit three dimensional
Schwarzian derivative is derived from a three dimensional PDE while $n \ (n\geq
4)$ dimensional $m\ (m\geq 4)$ order Schwarzian derivatives which are invariant
under the M\"obious transformation can be expressed by
means of the lower dimensional lower order Schwarzian derivatives.

Using two dimensional and three dimensional Schwarzian derivatives, we may
obtain various new Painlev\'e integrable models. Some of high dimensional
Painlev\'e integrable KdV type Schwarzian equations are given.

Though the Painlev\'e property is considered as a sufficient
condition of the integrability, and many of other integrable
properties like the Lax pair, symmetries, B\"acklund-Darboux
transformation and multi-soliton solutions can be obtained from
the usual Painlev\'e analysis for (1+1)- and (2+1)-dimensional models, it is
still open how to obtain other integrable properties from the
Painlev\'e analysis in higher dimensions especially in
(3+1)-dimensions. The more about the high dimensional Schwarzian
derivatives and the related high dimensional integrable models
is worthy of studying further.

\vskip.2in

The work was supported by
the Outstanding Youth
 Foundation and the National Natural Science Foundation
of China
 (Grant. No. 19925522), the Research Fund for the Doctoral Program
of Higher Education of China (Grant. No. 2000024832) and the
 Natural Science
Foundation of Zhejiang Province, China. The author
 is in debt to thanks the
helpful discussions with the Professor
 G-x Huang and Professor Q-p Liu.

\end{document}